\documentstyle[preprint,aps]{revtex}

\begin{document}
\draft
\title{Absolute measurement of the $5/2^+$ resonance of 
  $^{36}$Ar(p,$\gamma$)$^{37}$K at $E_{\rm{p}} = 918~{\rm{keV}}$}
\author{P.~Mohr$^{1,2}$, H.~Oberhummer$^{2}$, Gy.~Gy\"urky$^{3}$, 
  E.~Somorjai$^{3}$, \'A.Z.~Kiss$^{3}$, and I.~Borb\'ely-Kiss$^{3}$}
\address{
        $^1$ Institut f\"ur Kernphysik, Technische Universit\"at Darmstadt,
        Schlossgartenstr.~9, D--64289 Darmstadt, Germany \\
        $^2$ Institut f\"ur Kernphysik, Technische Universit\"at Wien,
        Wiedner Hauptstr.~8-10, A--1040 Vienna, Austria \\
        $^3$ Institute of Nuclear Research of the Hungarian Academy 
        of Sciences, P.O.~51, H--4001 Debrecen, Hungary
}
\date{\today}
\maketitle
\begin{abstract}
The strength of the 
$5/2^+$ resonance of the reaction
$^{36}$Ar(p,$\gamma$)$^{37}$K 
at $E_{\rm{p}} = 918~{\rm{keV}}$
was measured absolutely:
$\omega \gamma = (258 \pm 24)~{\rm{meV}}$. 
The new result is consistent with previous results.
This resonance was used in several experiments
as a calibration standard. 
Therefore we discuss our experimental uncertainties in detail
which is important for
the reliability of 
previous experiments and our new result.
\end{abstract}


\narrowtext

Recently, two resonances of the reaction 
$^{36}$Ar(p,$\gamma$)$^{37}$K 
have been measured for different purposes. 
The low-energy resonance
at $E_{\rm{p}} = 321~{\rm{keV}}$
is important for nuclear astrophysics because this resonance
determines the reaction rate of the
$^{36}$Ar(p,$\gamma$)$^{37}$K 
reaction in hot and explosive hydrogen burning \cite{il},
and the measurement of the properties of the resonance 
at $E_{\rm{p}} = 1422~{\rm{keV}}$
helped to resolve a discrepancy found for Gamow-Teller strength
distributions in $A = 37$ nuclei
which were determined from $\beta$ decay studies and from
(p,n) cross section data \cite{il2}.
However, all these resonance strengths have been determined relative
to the $5/2^+$ resonance
at $E_{\rm{p}} = 918~{\rm{keV}}$ 
(corresponding to $E_x = 2750.3~{\rm{keV}}$ in $^{37}$K)
which was measured absolutely by Goosman and Kavanagh 
in 1967 \cite{goos67,goos67a}
using a gas cell filled with enriched $^{36}$Ar and a NaI(Tl) detector
in close geometry:
$\omega \gamma = (208 \pm 30)~{\rm{meV}}$.
Unfortunately, no information on the uncertainties of about 15\%
is given in Ref.~\cite{goos67}.
In a recent work
de Esch and Van der Leun \cite{esch88} used implanted $^{36}$Ar targets
and germanium detectors to determine spins, parities, branching ratios,
and lifetimes of excited states in $^{37}$K, but they did not try to
determine absolute resonance strengths. Very recently, 
Hinnefeld {\it et al.} \cite{hinn98} have determined the resonance strength
indirectly. They measured the total width of the resonance 
$\Gamma = (0.94 \pm 0.67)~{\rm{eV}}$
by the Doppler shift attenuation method,
and they combined 
the total width $\Gamma$ and the recently measured branching ratio
$\Gamma_\gamma/\Gamma_{\rm{p}} = 0.69 \pm 0.15$ \cite{gar95}
to derive the resonance strength of 
$\omega \gamma = (680 \pm 490)~{\rm{meV}}$.
In this paper we present a direct and absolute measurement of 
the strength of the $5/2^+$ calibration resonance 
at $E_{\rm{p}} = 918~{\rm{keV}}$.

The experiment was performed at the Van-de-Graaff accelerator,
ATOMKI, Debrecen. The $^{36}$Ar targets consisted of implanted $^{36}$Ar 
into tantalum backings with a thickness of 0.4 mm. 
The targets were produced at the isotope separator 
of the University of Helsinki with an implantation energy
of $E_{\rm{imp}} = 20~{\rm{keV}}$ and a charge density of 
$\rho_{\rm{imp}} = 11~{\rm{mC/cm^2}}$
leading to an average implantation depth of 8.6 nm.

The number of implanted argon atoms was measured after the
(p,$\gamma$) experiment 
by proton-induced X-ray emission (PIXE) at the PIXE setup 
of ATOMKI \cite{PIXE}.
The X-rays were detected using a Si(Li) detector placed
at an angle of 135$^\circ$ relative to the proton beam axis
and covering a solid angle of $2.24 \times 10^{-4}$. 
The number of implanted $^{36}$Ar atoms was determined from 
the comparison of measurements on the Ta backing 
with and without implanted $^{36}$Ar atoms: 
$\rho(^{36}{\rm{Ar}}) = (1.33 \pm 0.09)~{\rm{\mu g/cm^2}}$.
The X-ray spectra were evaluated with the PIXYKLM code \cite{PIXE2}, which 
calculates the concentration and its uncertainty for the composite elements 
in ${\rm{\mu g/cm^2}}$.
The given overall uncertainty contains the uncertainties of 
beam current integration, detection solid angle, 
spectrum fitting procedure,
and X-ray absorption in the Ar/Ta target. 
The efficiency determination of the Si(Li) detector
was reported in Ref.~\cite{pal80},
and the applicability of the PIXE method to thick targets
was analyzed in Ref.~\cite{sza89}.
The result for the target thickness
is consistent with detailed
analyses of the implantation of argon atoms in tantalum 
\cite{sel67,alm61}.

The $\gamma$ rays from the $^{36}$Ar(p,$\gamma$)$^{37}$K reaction
were detected in a high-purity Germanium (HPGe) detector
with 40\% relative efficiency. The absolute efficiency of the HPGe detector
at $E_{\gamma} = 1332.5~{\rm{keV}}$
was determined from a calibrated $^{60}$Co source
with a quoted uncertainty of 0.7\% (National Office of Measures, Hungary), 
and the
energy dependence of the efficiency was measured using an uncalibrated
$^{56}$Co source which was produced at the cyclotron of ATOMKI.
The HPGe detector was placed at an angle of $\theta_\gamma = 55^\circ$ 
relative to the proton beam axis. At this angle the $P_2(\cos{\theta})$
contribution of the $\gamma$ ray angular distribution
vanishes, and the contribution of $P_4(\cos{\theta})$ is
negligible \cite{goos67,esch88}.
The $5/2^+$ resonance at
$E_{\rm{p}} = 918~{\rm{keV}}$
mainly decays to the ground state of $^{37}$K
with a $\gamma$ ray energy of
$E_{\gamma} = 2750.3~{\rm{keV}}$ \cite{il,goos67,goos67a,esch88}.
A typical $\gamma$ ray spectrum in the
$E_{\rm{p}} = 918~{\rm{keV}}$
resonance is shown in Fig.~\ref{fig:spect}.

The tantalum backing was thick enough to stop the protons. Therefore
the target was water-cooled to reduce the target deterioration during
the experiment. A cooling trap was installed to minimize the carbon
deposition on the implanted targets.
The accumulated charge was measured with an estimated accuracy of $\pm 3\%$
by standard charge integration. A suppression voltage of 
$-300~{\rm{V}}$ was used to reduce the influence of secondary electron
emission from the target. Typical beam currents were in the order
of $5~{\rm{\mu A}}$.

The resonance was measured three times. In the first step the
shape of the yield curve was measured in a relatively close
geometry (distance between detector and target: 
$d_{\rm{close}} = 26.8~{\rm{mm}}$). 
In the second step we measured
the $\gamma$ ray yield in the maximum of the yield curve at a larger
distance between detector and target 
($d_{\rm{far}} = 66.8~{\rm{mm}}$)
with good statistics.
At this distance $d_{\rm{far}}$,
the detection efficiency does not depend as sensitively 
on the geometry as at the close distance $d_{\rm{close}}$.
Finally, we moved back to the close distance
and checked again the shape and position of the yield curve.
No significant target deterioration was found during the experiment:
both, the energy and the width of the yield curve
could be measured again within their uncertainties.
The yield curve is shown in Fig.~\ref{fig:yield}.

The resonance strength was determined from two different procedures:
(i) from the shape and the amplitude of the yield curve and
(ii) from the maximum of the yield curve.

The resonance strength of a narrow resonance is given by the integration
of the yield curve over the resonance \cite{fow67}:
\begin{equation}
\omega \gamma \; = \;
\frac{\mu E_{\rm{r,c.m.}}}{(\pi \hbar)^2} \;
\frac{\int Y(E_{\rm{c.m.}}) \; dE_{\rm{c.m.}}}{N_{\rm{T}} \cal{E}} \quad ,
\label{eq:ref}
\end{equation}
where $\mu$ is the reduced mass, $Y$ the $\gamma$-ray yield per 
incident proton, $N_T$ the number of target atoms per ${\rm{cm^2}}$,
and $\cal{E}$ the efficiency of the HPGe detector.
The result for the resonance strength is
$\omega \gamma = 254 \pm 25~{\rm{meV}}$.
This result is independent of the stopping power of the protons in
the mixed Ar/Ta target.

From the one data point at the maximum of
yield curve at the distance $d_{\rm{far}}$
the resonance strength was determined using
the thick target yield formula with additional semi-thick
target corrections:
\begin{equation}
\omega\gamma = \frac{2 \epsilon}{\lambda_R^2} \cdot \frac{A_T}{A_P+A_T} 
\cdot Y_{\rm{thick}} \quad ,
\label{eq:thick_yield}
\end{equation}
where $\lambda_R$ is the proton wavelength at the resonance energy,
$A_P$ and $A_T$ the mass numbers of projectile and target,
$Y_{\rm{thick}}$ the thick target yield of the resonance, and
$\epsilon$ the effective stopping power of the ArTa compound:
\begin{equation}
\epsilon = \epsilon_{\rm Ar} + \frac{N_{\rm Ta}}{N_{\rm Ar}} \cdot
\epsilon_{\rm Ta} \quad .
\label{eq:stop}
\end{equation}
The semi-thick target correction was made by
\begin{equation}
Y(\xi)= Y_{\rm thick} \cdot \frac{2}{\pi} \cdot 
\arctan{\Big(\frac{\xi}{\Gamma'}\Big)} \quad ,
\end{equation}
where $\xi$ is the target thickness and $\Gamma'$ the energy difference 
between the quarter and three quarter maximum of the leading edge 
of the measured yield curve. This procedure leads to a resonance
strength of
$\omega \gamma = 270 \pm 40~{\rm{meV}}$.

Our final result for the resonance strength is the weighted average
of our two determinations:
$\omega \gamma = (258 \pm 24)~{\rm{meV}}$,
taking into account that the uncertainties
of the two determinations are independent 
with the exception of
the PIXE result and the strength of the calibrated $^{60}$Co source.
This result
is slightly higher than the
previous result 
$\omega \gamma = (208 \pm 30)~{\rm{meV}}$
of Goosman and Kavanagh \cite{goos67}.
Our result already includes a small correction 
due to the weak branching (1.4\%)
of this $5/2^+$ resonance to the $7/2^-$ excited state 
at $E_x = 1380~{\rm{keV}}$
in $^{37}$K \cite{il2,goos67a,esch88}.

The main uncertainties of our experiment are 
the number of Ar atoms in the target, 
the effective stopping power of the mixed Ar/Ta target (only in the
second determination of the resonance strength), 
and the efficiency of the HPGe detector.
The number of $^{36}$Ar atoms was measured by PIXE with an accuracy of
6.5\% (see above). 
The stopping power is dominated by the stopping of tantalum
because (i) $Z({\rm{Ta}}) = 73 >> Z({\rm{Ar}}) = 18$, and
(ii) the number of Ta atoms is a factor of $8.3~\pm~10\%$ higher than the
number of Ar atoms. The stopping power of tantalum was taken from the 
Ziegler tables \cite{ziegler}, and the value was corrected by 6\% 
because of the additional stopping power of argon. The overall uncertainty of
the effective stopping power (including the argon contribution)
was estimated as 11\%.
The relative detector efficiency was determined from many
$\gamma$ ray lines from the $^{56}$Co source with energies above and
below the relevant $\gamma$ ray energy of $E_{\gamma} = 2750.3~{\rm{keV}}$.
The uncertainty of the efficiency ${\cal{E}}(2750.3~{\rm{keV}})$ is given by
the uncertainties of the parameters $a$ and $b$ in the following equation:
\begin{equation}
\log{{\cal{E}}} = a \cdot \log{E_\gamma} + b
\label{eq:eff}
\end{equation}
For our detector in the far geometry we obtained
${\cal{E}}_{\rm{far}}(2750.3~{\rm{keV}}) = 1.92 \cdot 10^{-3}\pm 5.5\%$,
and in the close geometry
${\cal{E}}_{\rm{close}}(2750.3~{\rm{keV}}) = 5.39 \cdot 10^{-3}\pm 6.7\%$,
including the minor uncertainty from the calibrated $^{60}$Co source.
Because of the relatively small distances $d_{\rm{close}}$ and $d_{\rm{far}}$
cascade summing effects from coincident $\gamma$ rays in the decay schemes of
$^{56}$Co and $^{60}$Co had to be taken into account.
The total uncertainty of each of 
our two results is given by the quadratic sum of
the above uncertainties.
Finally, after averaging both results, we end up with an overall uncertainty
of 9.5\% which is slightly smaller than the 14.4\% uncertainty
of Ref.~\cite{goos67} which was not discussed at all.

In conclusion, we have measured the resonance strength of the 
$5/2^+$ resonance of the reaction
$^{36}$Ar(p,$\gamma$)$^{37}$K 
at $E_{\rm{p}} = 918~{\rm{keV}}$
absolutely.
The average of our new result
$\omega \gamma = (258 \pm 24)~{\rm{meV}}$ and
the previously used strength of Goosman and Kavanagh 
$\omega \gamma = (208 \pm 30)~{\rm{meV}}$ \cite{goos67}
gives
$(\omega \gamma)_{\rm{average}} = (238 \pm 19)~{\rm{meV}}$
which is consistent with the results of both experiments
within their uncertainties,
and the overall uncertainty of the resonance strength
is reduced from about 15\% to less than 10\%.
We did not include the result of Ref.~\cite{hinn98}
into the average value
because of their large (72\%) uncertainties.
 
The experimental results of Refs.~\cite{il,il2} should be normalized to the
new resonance strength of the $5/2^+$ calibration resonance. 
The final results are summarized in Table \ref{tab:tab1}.
We obtain
for the resonance strength
of the resonance at 321~keV $\omega \gamma = (7.0 \pm 1.0) \times
10^{-4}~{\rm{eV}}$ instead of $\omega \gamma = (6.1 \pm 1.4) \times
10^{-4}~{\rm{eV}}$~\cite{il},
and for the resonance at 1422~keV $\omega \gamma = (6.9 \pm 1.0) \times
10^{-4}~{\rm{eV}}$ instead of $\omega \gamma = (6.0 \pm 1.5) \times
10^{-4}~{\rm{eV}}$~\cite{il2}.
However, the main conclusions of Refs.~\cite{il,il2,hinn98} 
remain unchanged and are confirmed by this experiment.

\acknowledgements 
One of us (P.~M.) wants to thank for the very kind hospitality during
the experiment at ATOMKI. A special thank to P. Tikkanen and M. Wiescher for 
providing the targets for this measurement. This work was supported by 
Austrian-Hungarian Exchange Program (project A-20/96),
Fonds zur F\"orderung der
Wissenschaftlichen Forschung in \"Osterreich (project S7307--AST),
Deutsche Forschungsgemeinschaft (DFG) (project Mo739/1), and OTKA 
(No. T016638 and CW015604).

\begin{table}
\caption{\label{tab:tab1} 
Propoerties of $^{36}$Ar(p,$\gamma$)$^{37}$K resonances. The partial widths
of the 918 keV resonance are derived from the adopted resonance strength
and the recently measured branching ratio 
$\Gamma_\gamma/\Gamma_{\rm{p}} = 0.545 \pm 0.030$ (weighted average from
\protect\cite{tri95,gar95,mag95}). 
The total width $\Gamma = 348 \pm 28~{\rm{meV}}$
corresponds to a lifetime of $\tau = 1.89 \pm 0.15~{\rm{fs}}$
which is in agreement with 
the upper limit $\tau < 3~{\rm{fs}}$ \protect\cite{esch88}
and in rough agreement with $\tau = 0.7 \pm 0.5~{\rm{fs}}$
\protect\cite{hinn98}.
}
\begin{center}
\begin{tabular}{cccccccccc}
  $E_{\rm{p}}$ & $E_x$      & $J^\pi$ 
  & $\omega \gamma$\tablenotemark[1]
  & $\omega \gamma$\tablenotemark[2]
  & $\omega \gamma$
  & $\Gamma_{\rm{p}}$
  & $\Gamma_{\gamma}$
  & $\Gamma$ \\
  (keV) & (keV)      &  
  & (meV)
  & (meV)
  & (meV)
  & (meV)
  & (meV)
  & (meV) \\
\hline
  918 & 2750      & $5/2^+$ 
  & $208 \pm 30$  & $258 \pm 24$  & $238 \pm 19$\tablenotemark[3]
  & $225 \pm 20$  & $123 \pm 10$  & $348 \pm 28$ \\
  321 & 2170      & $3/2^-$ 
  & $0.61 \pm 0.14$ &             & $0.70 \pm 0.10$\tablenotemark[4]
  &               &               &              \\
  1422 & 3240      & $(5/2,7/2)^+$ 
  & $0.60 \pm 0.15$&              & $0.69 \pm 0.10$\tablenotemark[4] 
  &               &               &              \\
\end{tabular}
\tablenotetext[1]{from Refs.~\protect\cite{goos67,il,il2}}
\tablenotetext[2]{this work}
\tablenotetext[3]{weighted average from this work and 
Ref.~\protect\cite{goos67}}
\tablenotetext[4]{adopted strengths: 
from Refs.~\protect\cite{il,il2} normalized to the
new adopted strength of the 918 keV resonance given in the first line.}
\end{center}
\end{table}

\begin{figure}
\caption{
        \label{fig:spect} 
        Typical $\gamma$ ray spectrum of the 
        $E_{\rm{p}} = 918~{\rm{keV}}$
        resonance in the reaction
        $^{36}$Ar(p,$\gamma$)$^{37}$K.
        Note the logarithmic scale of the full spectrum and the
        linear scale of the inset at the relevant transition energy
        $E_{\gamma} = 2750.3~{\rm{keV}}$.
}
\end{figure}

\begin{figure}
\caption{
        \label{fig:yield} 
        Yield curve 
        of the reaction
        $^{36}$Ar(p,$\gamma$)$^{37}$K
        measured at the
        $E_{\rm{p}} = 918~{\rm{keV}}$
        resonance.
        }
\end{figure}

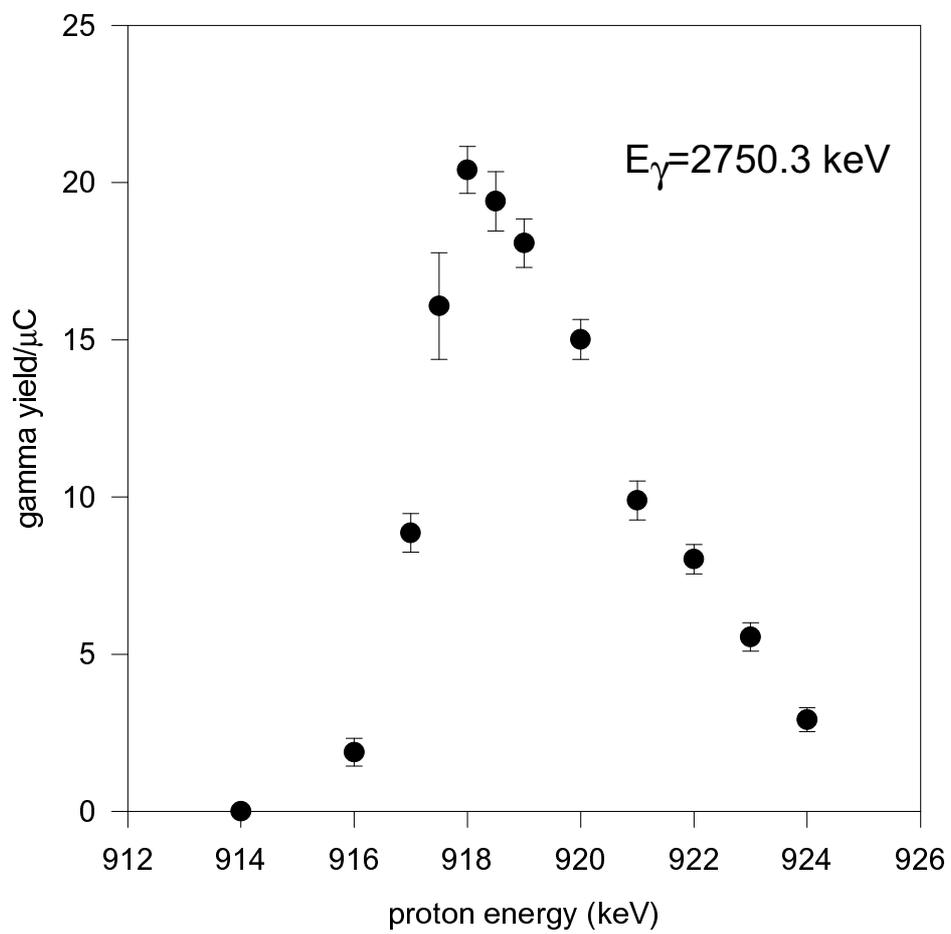
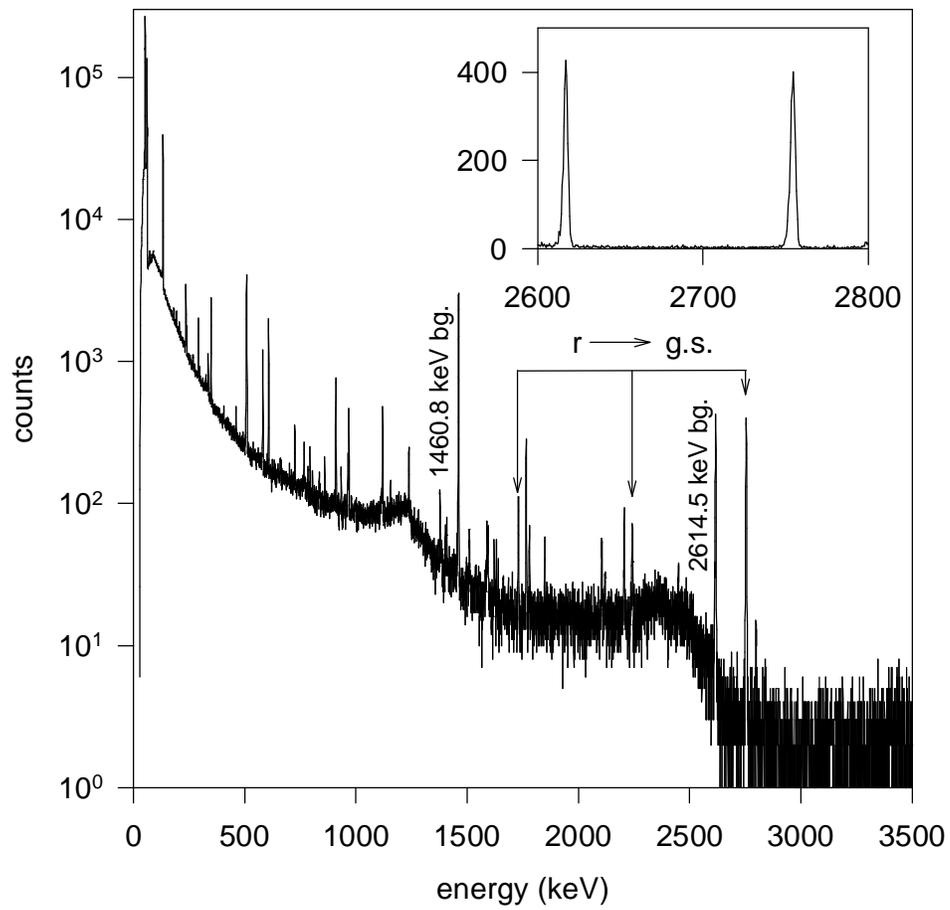
\end{document}